\documentclass[twocolumn]{jpsj2}

\def\cuclbr{CuCl$_{2x}$Br$_{2(1-x)}$($\gamma$-pic)$_2$ }

\def\cucl{CuCl$_{2}$($\gamma$-pic)$_2$ }
\def\const{\rm const}

\def\eff{\rm eff}
\def\simleq{ \mbox{ \raisebox{-1.0ex}{ $\stackrel{<}{\sim}$ } } }
\def\simgeq{ \mbox{ \raisebox{-1.0ex}{ $\stackrel{>}{\sim}$ } } }
\def\ln{\mbox{ln}}

\def\v#1{\mib #1}
\def\clcl{${\mbox{Cu}}<^{\mbox{Cl}}_{\mbox{Cl}}>{\mbox{Cu}}$ }
\def\brbr{${\mbox{Cu}}<^{\mbox{Br}}_{\mbox{Br}}>{\mbox{Cu}}$ }
\def\clbr{${\mbox{Cu}}<^{\mbox{Br}}_{\mbox{Cl}}>{\mbox{Cu}}$ }
\def\runtitle{Mixture of Bond-Alternating and Uniform $S=1/2$  Antiferromagnetic Heisenberg Chains
}

\def\runauthor{Kazuo {\sc Hida}}

\title
{
Ground State and Magnetization Process of the Mixture of Bond-Alternating and Uniform $S=1/2$ Antiferromagnetic Heisenberg Chains
}

\author
{Kazuo {\sc Hida}
\footnote{e-mail: hida@phy.saitama-u.ac.jp}}

\inst
{
Department of Physics, Faculty of Science,\\ Saitama University, Saitama 338-8570
}

\recdate
{
\today
 }

\abst
{
The mixture of bond-alternating and uniform  $S=1/2$  antiferromagnetic Heisenberg chains is investigated by the density matrix renormalization group method. The ground state magnetization curve is calculated and the exchange parameters are determined by fitting to the experimentally measured magnetization curve of \cuclbr . The low field behavior of the magnetization curve and low temperature behavior of the magnetic susceptibility are found to be sensitive to whether the bond-alternation pattern (parity) is fixed all over the sample or randomly distributed. The both quantities are compatible with the numerical results for the random parity model.
}
\kword
{
\cuclbr, random Heisenberg chain, density matrix renormalization group, magnetization process, random singlet phase, quantum Griffiths phase
}
\begin{document}
\maketitle

\section{Introduction}
In recent studies of quantum many body problem, the ground state properties of the random quantum systems have been attracting a renewed interest. Among them, the effect of randomness in one-dimensional quantum spin systems has been extensively studied theoretically and experimentally\cite{dsm1,dsm2,df1,kh1,hy1,hy2,kh2,uchiprog,uchi2,azuma,fuku,ajiro,waki}.

The real space renormalization group (RSRG) analysis of the $S=1/2$ random bond antiferromagnetic Heisenberg chain\cite{dsm1,dsm2,df1} has shown that the ground state of the $S=1/2$ uniform antiferromagnetic Heisenberg chain is unstable against the bond randomness of infinitesimal strength and the random singlet phase is realized in which the spins form singlet pairs randomly with distant partners.\cite{df1} In this phase, the spin-spin correlation decays by the power law with exponent $-2$ and log-averaged energy gap scales as $<\ln \Delta> \simeq -\alpha\sqrt{N}+\const.$, where $N$ is the system size and $\alpha$ is a constant. This has been confirmed numerically\cite{kh1} by means of the density matrix renormalization group (DMRG) method\cite{wh1}.

On the other hand, the effect of randomness on the spin gap state of one-dimensional quantum spin systems has been extensively studied\cite{hy1,hy2,kh2,uchiprog,uchi2,azuma,fuku} related with the doping effect in bond-alternating chains\cite{uchiprog,uchi2} and spin ladders\cite{azuma}. Hyman and coworkers\cite{hy1,hy2} have applied the RSRG method to the spin-1/2 random dimerized antiferromagnetic Heisenberg chain and have shown that the dimerization is a releveant perturbation to the random singlet phase\cite{hy1,hy2}. This implies that the ground state of the spin-1/2 dimerized antiferromagnetic Heisenberg chain is stable against infinitesimal randomness. They also argued that the ground state of this model is the random dimer phase which belongs to the quantum Griffiths phase in which the spin-spin correlation length is finite and log-averaged energy gap scales as $<\ln \Delta> \sim -z\ln{N}+\const.$ with finite dynamical exponent $z$. This is also verified by the DMRG calculation by the present author\cite{kh2}.

Motivated by the quite distinct nature of the uniform chain and bond-alternating chain against randomness, Ajiro and coworkers experimentally investigated the magnetic properties of the compound \cuclbr \cite{ajiro,waki} which is  the mixed compound of $S=1/2$ uniform and bond-alternating antiferromagnetic Heisenberg chains. The $x=1$ compound CuCl$_2$($\gamma$-pic)$_2$ is an $S=1/2$ bond-alternating antiferromagnetic Heisenberg chain\cite{groot} and the $x=0$ compound  CuBr$_2$($\gamma$-pic)$_2$ is an  $S=1/2$ uniform  antiferromagnetic Heisenberg chain\cite{ajiro}. In order to understand the low temperature magnetic properties of these compounds, we theoretically investigate the ground state and low energy properties of the mixture of bond-alternating and uniform $S=1/2$ antiferromagnetic Heisenberg chains using  DMRG method in the present work.

This paper is organized as follows. In the next section, the thoretical model which describes the mixed compound \cuclbr is presented. In section 3, the magnetization curve is calculated and the exchange parameters are determined from overall behavior of the magnetization curve. The numerical results of the low energy magnetic excitation spectrum is presented in section 4. Based on the finite size scaling analysis of the log-averaged energy gap, the low temperature singularity of the magnetic suscsptibility is predicted and the explanation of the experimentally observed temperature dependence of susceptibility  data is given. The last section is devoted to summary and discussion.

\section{Theoretical Model of \cuclbr}

In the compound \cuclbr, the Cu-Cu bond is bibridged by Cl and/or Br ions, so that there exist \clcl, \clbr and \brbr bonds in this compound.  In the pure compound \cucl, two kinds of \clcl bonds alternate along the chain, although this material appears to be a uniform chain from chemical formula. This bond alternation is induced by the freezing transition of rotational motion of methyl-group at 50K\cite{groot}. For finite Br-concentration $(0 < x < 1)$, this compound can be regarded as an assembly of the finite length clusters connected only by the \clcl bonds (hereafter called 'bond-alternating cluster'; abbrivaiated as BAC) with \brbr and/or \clbr bonds in between. However, it is not obvious whether the rotational order of methyl-group remains long ranged even in the mixed compound. If this order is long ranged, the bond alternation pattern is common among different BAC's even though they are separated by \brbr or \clbr bonds. In what follows, this case is called  {\it fixed parity} case. On the other hand, if the rotational order of methyl-group is also cut into short range order by Br-substitution on Cl sites, the bond alternation patterns of the \clcl bonds are uncorrelated among different BAC's once they are separated by \clbr or \brbr bonds. This case is called {\it random parity} case. For the moment, there is no experimental evidence which case is realized in \cuclbr. In the following, we therefore investigate both cases theoretically.

 The strength of these bonds are denoted by  $J$ (strong \clcl bond), $\alpha J$ (weak \clcl bond), $J'$ (\clbr bond) and $J''$ (\brbr bond), respectively in the following Hamiltonian which describes the mixed chain. 

\begin{equation}
H =  \sum_{i=1}^{N}2J_i \v{S}_{i} \v{S}_{i+1},
\end{equation}
\begin{equation}
J_i = \left\{\begin{array}{ll}
\left.\begin{array}{ll}J &\mbox{for} \ i+i_p=\mbox{even} \\
\alpha J &  \mbox{for} \ i+i_p=\mbox{odd}
\end{array}\right\}  & p= x^2, \\
\ J'   & p=2x(1-x), \\
\ J''  & p=(1-x)^2,
\end{array}\right.
\end{equation}
where  $p$ is the probability of realization of each type of bonds.  We define the integer 'parity' $i_p$ for each BAC. Corresponding to the fixed and random parity cases explained above, the parity $i_p$ takes the following values in each cases.
\begin{enumerate}
\item  Random parity case : $i_p$ takes the values 0 or 1 randomly in different BAC's. 
\item  Fixed parity case : $i_p=0$ for all BAC's. 
\end{enumerate}

\begin{figure}
\centerline{\includegraphics[width=80mm]{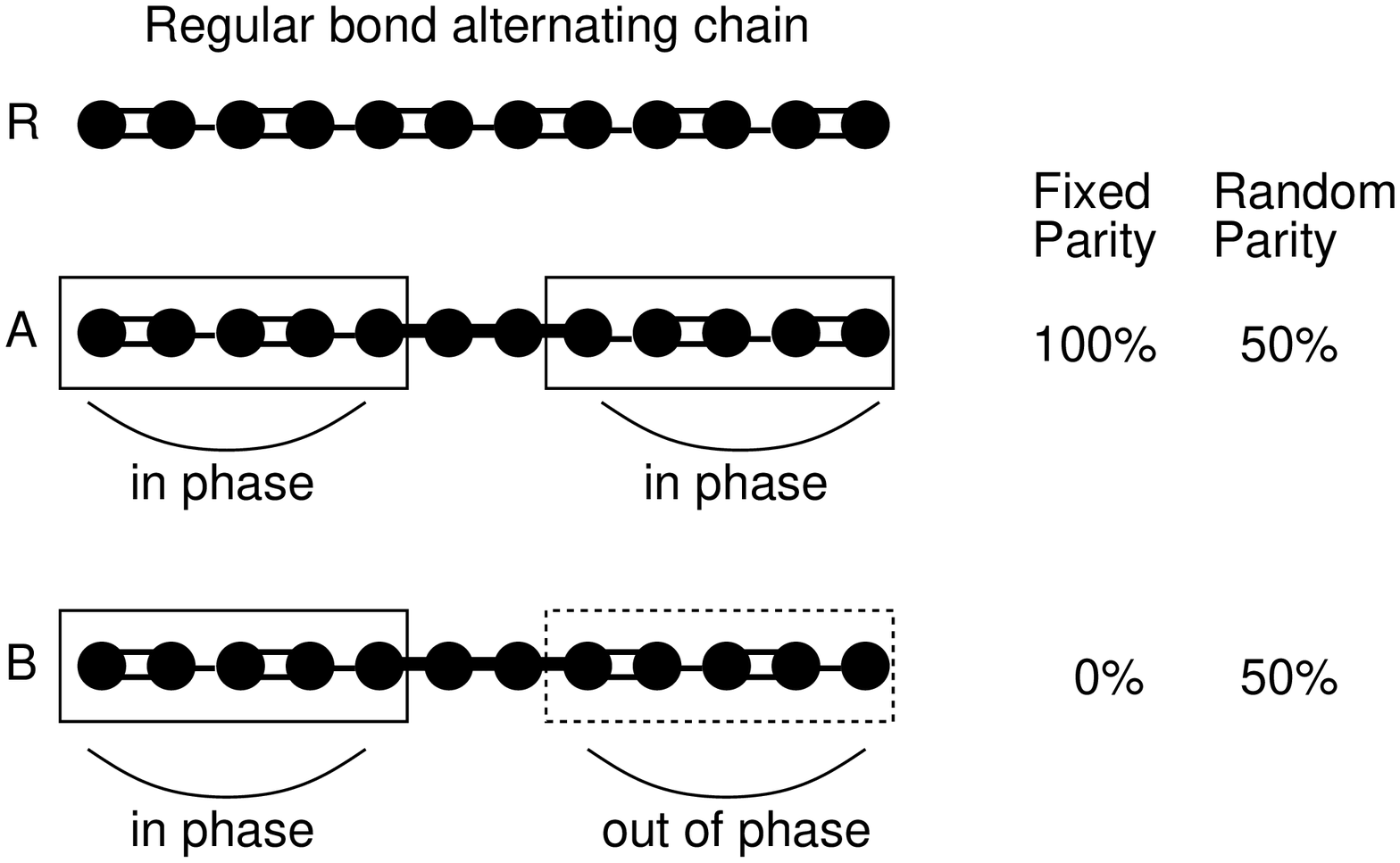}}
\caption{Schematic view of the local bond configuration. The thick lines denote the \clbr or \brbr bonds and the thin single and double lines, the alternating \clcl bonds. The uppermost configuration R corrsponds to the regular bond-alternating chain (\cucl). Two possible local bond configurations in the mixed chain \cuclbr are shown as A and B. The clusters in the rectangles are bond-alternating clusters (BAC's). In the random parity case, the type-A and type-B configurations occur with the same weight. In the fixed parity case, only type-A configuration occurs. }
\label{cluster}
\end{figure}

The possible local bond configurations realized in each case can be visually explained using Fig. \ref{cluster}. The uppermost configuration R shows the regular bond-alternating chain corresponding to \cucl. The local configurations of type-A and B can occur in the mixed chain \cuclbr. The two BAC's in type-A configuration are in phase with regular chain R. On the other hand, in type-B configuration, the left BAC in the solid rectangle is in phase with R while the right BAC in the dotted rectangle is out of phase from R. The local configurations of type-A and B (with its mirror inversion) occur with the same weight in the random parity case but only type-A configuration occurs in the fixed parity case.

\section{Magnetizaton Process}
\begin{figure}
\centerline{\includegraphics[width=80mm]{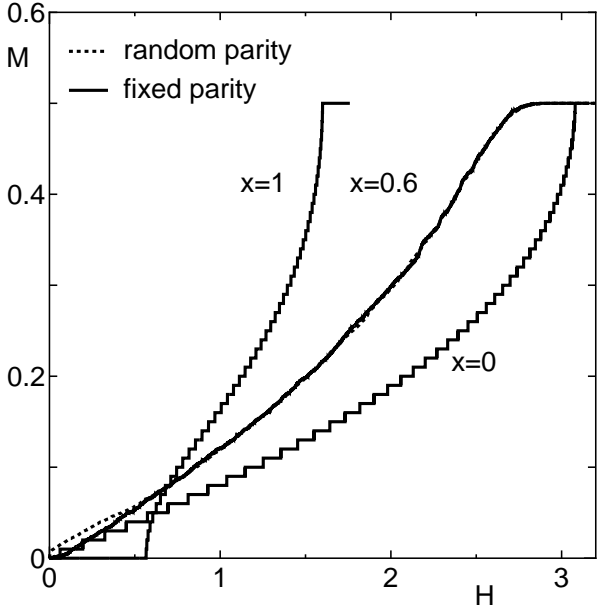}}
\caption{Magnetization curves of the fixed parity model (solid line) and random parity model (dotted line) with $x=0.6$. The stepwise lines are the magnetization processes of pure systems ($x=0$ and $x=1$). For $x=0.6$, the error bars are comparable with the width of the lines.}
\label{magcurve}
\end{figure}

From the measurement for the pure systems ($x=0, 1$), the coupling constants $J, \alpha$ and $J''$ are determined as,\cite{ajiro,waki}
\begin{equation}
J= 13.2 {\rm K},\ J''= 20.3 {\rm K},\ \alpha=0.6.
\end{equation}
Using these parameters, we have calculated the magnetization curve of the mixed chain using DMRG method for various choices of $J'$ and determined $J'/J= 1.3$ so as to reproduce the overall magnetization curve of \cuclbr. The chain length $N$ is fixed to 100 and the average is taken over 100 samples. The maximum number of the states kept in each subsystem in the course of DMRG calculation is 300. It should be noted that the overall magnetization curve does not depend whether the parity is fixed or random, except for the low field part where the magnetization curve is more convex in the random parity model as shown in Fig. \ref{magcurve}.
The low field part of the magnetization curve reported in refs. \citen{ajiro} and \citen{waki} seems to be consistent with the random parity model. However, due to the finite temperature effect in the experimental data and finite size effect in the numerical calculation, the direct quantitative comparison is difficult. Actually, the most significant difference between the two models is the low field singularity of the magnetization determined by the low energy singularity of the magnetic excitation spectrum. At finite temperatures, however, this singularity is rounded to yield finite magnetic susceptibility which diverges or vanishes as the temperature is lowered. Therefore, for the quantitative comparison with experiment, we concentrate on the low temperature singularity of the susceptibility which can be deduced from the finite size scaling analysis of the low energy magnetic excitation spectrum in the next section.

\section{Low energy magnetic excitation spectrum}

\subsection{Fixed parity case}

The lowest excitation gaps with total magnetization $S^{\rm tot}_z=1$ for each sample is calculated using the infinite size DMRG method\cite{kh1} for $10 \leq N \leq 80$ ($x \leq 0.5$) and for $10 \leq N \leq 160$ ($x \geq 0.6$). The average is taken over 200 samples. The number of the states kept in each subsystem during the course of the DMRG calculation is $60$. The accuracy is confirmed by checking the coincidence of the energy gap with $(S_{\rm tot}, S_{\rm tot}^z)=(1,0)$ and that with $(S_{\rm tot}, S_{\rm tot}^z)=(1,1)$ taking into account the $SU(2)$ symmetry of the present system.

\begin{figure}
\centerline{\includegraphics[width=80mm]{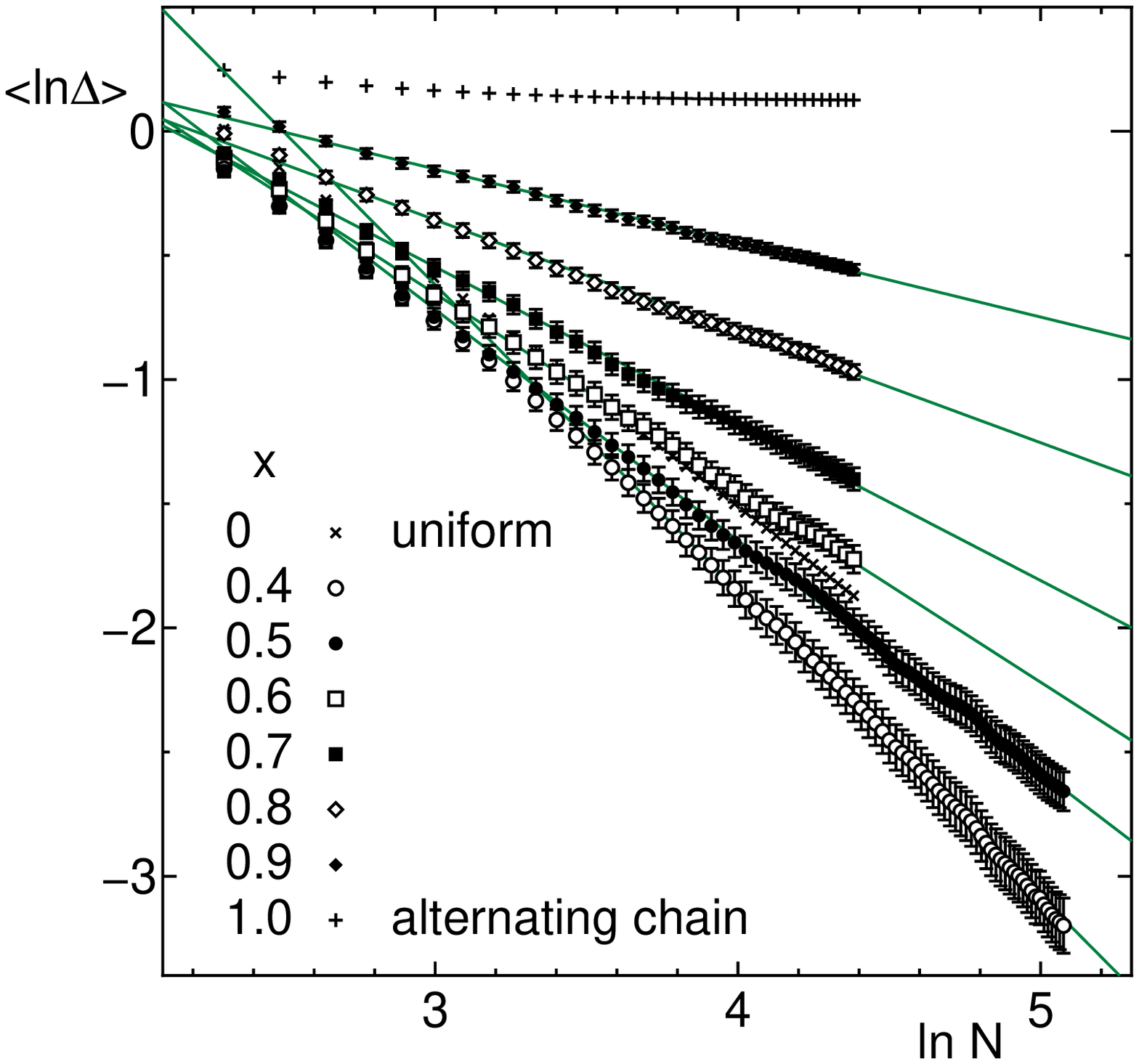}}
\caption{System size dpendence of  $<\ln \Delta>$  in the fixed parity case. The gap is measured in units of $J$.}
\label{fmag}
\end{figure}

The log-averaged energy gap $<\ln \Delta >$ is plotted against  $\ln N$ in Fig. \ref{fmag} which show a fairly good linear behavior 
\begin{equation}
<\ln \Delta > = \ln \Delta_0 - z \ln N ,
\label{fxfit}
\end{equation}
for large $N$ and $x \geq 0.4$. For $x \leq 0.3$, the data are not well fitted by (\ref{fxfit}). Presumably, in this regime, the concentration of the alternating bond $x^2$ is too small and true asymptotic behavior is not observable for the present system sizes. This result implies that the ground state of this model is the quantum Griffiths phase characterized by the finite dynamical exponent $z$ at least  for $ x \geq 0.4$. In the quantum Griffiths phase, the low temperature susceptibility $\chi(T)$ and low field magnetization at zero temperature $M(H)$ should behave as $\chi(T) \sim T^{1/z-1}$ and $M(H) \sim H^{1/z}$.\cite{hy1,hy2}

Physically, this result is reasonable. In this model, the dimerization pattern is fixed over the whole system. As pointed out by Hyman and coworkers\cite{hy1}, the dimerization is a relevant perturbation to the random singlet phase and drives the system into the random dimer phase which belongs to the quantum Griffiths phase. 

\begin{figure}
\centerline{\includegraphics[width=80mm]{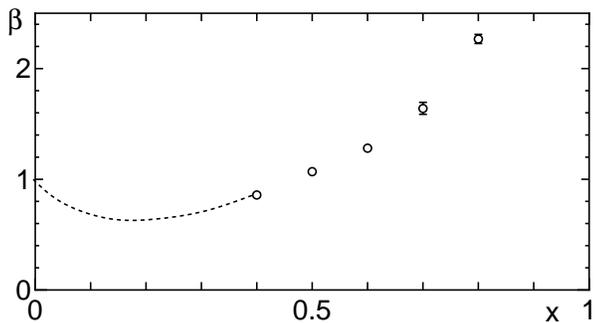}}
\caption{The exponent $\beta$ plotted against $x$ in the fixed parity case.}
\label{expo}
\end{figure}

In the experiment, it is found that the low temperature susceptibility of \cuclbr shows a divergent behavior as $\chi \sim T^{\beta-1}$ with an exponent $\beta \sim 0.67$\cite{ajiro,waki,ajipr}. The value of $\beta$ slightly depends on the concentration $x$. At first sight, the above result appears to be consistent with the prediction of the fixed parity model, if we identify $z=1/\beta$ as proposed in ref. \citen{ajiro}. Based on this idea, we determine the exponent $z$ by fitting $<\ln \Delta >$ in Fig. \ref{fmag} by Eq. (\ref{fxfit}). The result is shown in Fig. \ref{expo} in terms of $\beta =1/z$.  Although the value of $\beta$ for small $x$ is difficult to determine from the numerical data, we may expect it would roughly behave as shown by the dotted line considering that $\beta=1$ for $x=0$ (Luttinger liquid). The exponent $\beta$ is, however, always  larger than or around unity as shown in Fig. \ref{expo} within the regime $x \geq 0.4$. Therefore the fixed parity model does not explain the experimentally observed exponent $\beta \sim 0.67$. Actually, if we look into the experimental data closely, the exponent $\beta$ turns out to decrease with $x$, although the temperature of the experiment is not low enough to exclude the contribution of higher energy excitations. On the contrary,  in the fixed parity model, $\beta$ increases rapidly with $x$ for $x \simgeq 0.4$ as shown in Fig. \ref{expo}. The physical reason of the sharp increase of $\beta$ is obvious. In the limit $x \rightarrow 1$, the system is gapped, so that $z$ should tend to zero. 


\subsection{Random parity case}
In this case also, the lowest excitation gaps with total magnetization $S^{\rm tot}_z=1$ for each sample is calculated using the DMRG method for $N \leq 80$ in the same way as in the preceding subsection. The maximum number $m$ of the state kept in the course of DMRG calculation is $m=200$.

\begin{figure}
\centerline{\includegraphics[width=80mm]{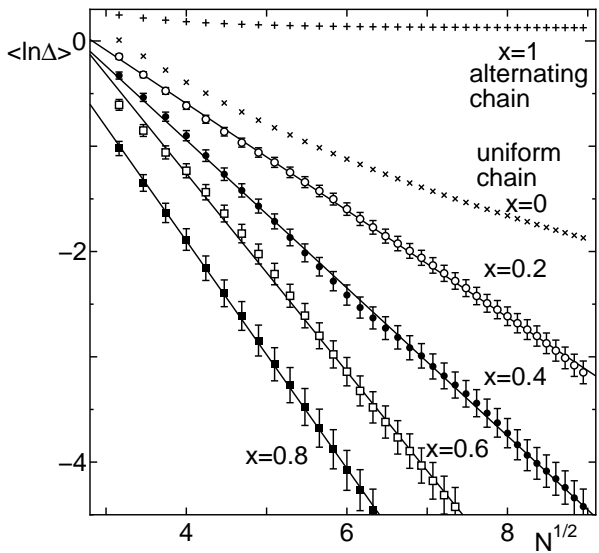}}
\caption{System size dependence of $<\ln \Delta>$ in the random parity case. The gap is measured in units of $J$.}
\label{rmag}
\end{figure}

The log-averaged magnetic energy gap $<\ln \Delta >$ is plotted against $\sqrt{N}$ in Fig. \ref{rmag} which show a fairly good linear behavior 
\begin{equation}
<\ln \Delta >=\ln \Delta_0-\alpha \sqrt{N}.
\label{rsgap}
\end{equation}
This implies that the ground state is the random singlet state in which the low temperature susceptibility $\chi(T)$ and low field magnetization at zero temperature $M(H)$ should behave as $\chi(T) \sim 1/(T(\ln T)^2)$ and $M(H) \sim (\ln(1/H))^{-2}$.\cite{df1} In this case, the dimerization pattern is not long ranged. Therefore the translational symmetry is preserved on average and the random singlet phase remains stable. 

However, it should be noted that this asymptotic form for $\chi$ is only valid in the true low temperature limit. At finite temperatures, we may define the temperature dependent effective value of $z$ even in the random parity case in the following way.

Let us define the size dependent effective value of $z$ by
\begin{equation}
z_{\eff} \equiv -\frac{d <\ln \Delta >}{d \ln N},
\end{equation}
based on the formula (\ref{fxfit}). This size dependent exponent can be interpreted as the energy dependent effective exponent because the typical energy scale of the cluster is related with the cluster size $N$ by the relation (\ref{rsgap}). Thus we find that $z_{\eff}$ is given by 
\begin{equation}
z_{\eff}=\frac{\alpha}{2}\sqrt{N}=\frac{1}{2}<\ln (\frac{\Delta_0}{\Delta})>.
\label{zene}
\end{equation}
Therefore, in the temperature range around $T$, we can define the temperature dependent effective value of $z$  by
\begin{equation}
z_{\eff}(T) = \frac{1}{2}\ln(\frac{\Delta_0}{k_{\rm B}T})
\label{ztemp}
\end{equation}
by setting $<\ln\Delta>=\ln (k_{\rm B}T)$ in (\ref{zene}). 

\begin{figure}
\centerline{\includegraphics[width=80mm]{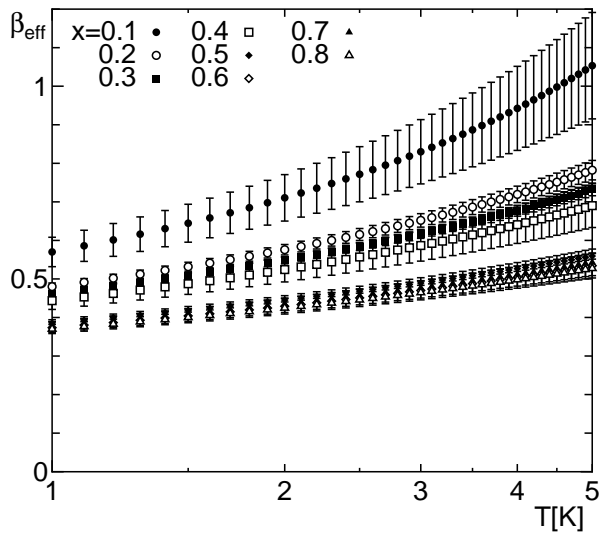}}
\caption{The effective exponent $\beta_{\eff}$ plotted against temperature $T$K. }
\label{zeffal}
\end{figure}

\begin{figure}
\centerline{\includegraphics[width=80mm]{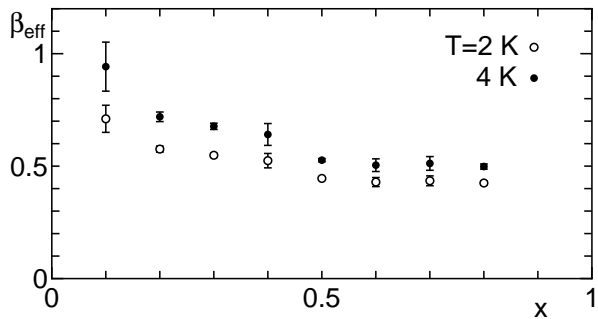}}
\caption{The effective exponent $\beta_{\eff}$ plotted against Cl concentration $x$.}
\label{zefftemp}
\end{figure}

From the fit to (\ref{rsgap}) we have determined the value of $\Delta_0$. Around the temperature range where the measurement is made ($T \simgeq 2 K $), the values of the effective exponent $\beta_{\eff}(T) \equiv 1/z_{\eff}(T)$ are plotted against $T$ in Fig. \ref{zeffal} for avrious values of $x$. They are also plotted against $x$ for $T=2$K and 4K in Fig. \ref{zefftemp}. It is clearly seen that  $\beta_{\eff}$ ranges from 0.4 to 0.7 for $0.8 \geq x \geq 0.2$ . This explains why the measured susceptibility exponent is around these values. Especially, at the fixed temperature, the exponent $\beta$ decreases with $x$ in accordance with the experimental tendency. 

Physically, the $x$-dependence of $\beta_{\eff}$ is understood as follows. The quantity $\Delta_0$ is essentially the bare energy scale of the system. In the spirit of RSRG method for random quantum spin chains\cite{dsm1,dsm2,df1}, it is of  the order of the largest exchange coupling $2J'' \sim 40.6$K which is decimated in the first step renormalization. Therefore we may roughly estimate as $\beta_{\eff} \sim 2/\ln (2J''/k_BT) \sim 0.66$ at $T=2$K and this value is insensitive to $x$. However, as $x$ appraoches zero, the ground state should tend to the Luttinger liquid characterized by $z=\beta=1$. Therefore $\beta_{\eff}$ increases to approach this value with the decrease of $x$. Actually, for small $x$, there exist many large uniform clusters coupled with the strongest bond $J''$ so that the first step renormalization cannot start with the single bond. Therefore, the effective value of $\Delta_0$ is reduced from $J''$ and $\beta_{\eff}$ is enhanced. 


\section{Summary}

The magnetization process of the mixture of uniform and alternating-bond  $S=1/2$ antiferromagnetic Heisenberg chain is calculated using the DMRG method in the ground state. The magnetization processes of random and fixed parity cases  are similar to each other except for the low field singularity which reflects the low energy spectrum. Comparing with the experimental magnetization curve  for \cuclbr the exchange parameters are determined. The low field magnetization curve of  \cuclbr looks similar to that of the random parity case although  the direct comparison is difficult due to the finite temperature effect in experiment and finite size effect in numerical calculation. 

For the quantitative analysis of the low field limit, the low energy energy spectrum obtained by the DMRG method is analyzed using the finite size scaling method. The ground state is the random singlet phase for the random parity model and it is the quantum Griffiths phase in the fixed parity model.  The low temperature singularity of the experimentally observed susceptibility of \cuclbr is consistent with the theoretical results for the random parity model assuming the temperature dependent effective exponent $\beta_{\rm eff}$. 

It should be noted that our calculation predicts that the effective value of $\beta$ decreases with the Cl concentration for $x \simleq 0.5$. Although this seems to be compatible with the experimental data, more precise measurement of susceptibility at lower temperatures is hoped in the future including the possibility of direct measurement of the random singlet  behavior $\chi(T) \sim 1/(T(\ln T)^2)$.

The author is grateful to Y. Ajiro for valuable discussion and for explaining the details of the analysis of the experimental data. The computation in this work has been done using the facilities of the Supercomputer Center, Institute for Solid State Physics, University of Tokyo and the Information Processing Center, Saitama University.  This work is supported by a Grant-in-Aid for Scientific Research from the Ministry of Education, Culture, Sports, Science and Technology, Japan.

\end{document}